\documentclass[%
 aip,
 amsmath,amssymb,
 reprint,%
]{revtex4-1}

\usepackage{graphicx}
\usepackage{dcolumn}
\usepackage{bm}

\usepackage[T1]{fontenc}
\usepackage{mathptmx}

\usepackage{graphicx}
\usepackage{caption}
\usepackage{amsmath}
\usepackage{amsthm}
\usepackage{amsfonts}
\usepackage{sidecap}
\usepackage{mathtools}
\usepackage{adjustbox}
\usepackage{siunitx}

\begin{document}

\preprint{AIP/123-QED}

\title[]{Defect Physics of Pseudo-cubic Mixed Halide Lead Perovskites from First Principles}


\author{Arun Mannodi-Kanakkithodi}
 \email{mannodiarun@anl.gov}
\affiliation{Center for Nanoscale Materials, Argonne National Laboratory, Argonne, Illinois 60439, USA}%

\author{Ji-Sang Park}
\affiliation{Department of Materials, Imperial College London, London SW7 2AZ, United Kingdom}%

\author{Alex B. F. Martinson}
\affiliation{Materials Science Division, Argonne National Laboratory, Argonne, IL 60439, USA}%

\author{Maria K.Y. Chan}%
 \email{mchan@anl.gov}
\affiliation{Center for Nanoscale Materials, Argonne National Laboratory, Argonne, Illinois 60439, USA}%

\date{\today}
\begin{abstract}
Owing to the increasing popularity of lead-based hybrid perovskites for photovoltaic (PV) applications, it is crucial to understand their defect physics and its influence on their optoelectronic properties. In this work, we simulate various point defects in pseudo-cubic structures of mixed iodide-bromide and bromide-chloride methylammonium lead perovskites with the general formula MAPbI$_{3-y}$Br$_{y}$ or MAPbBr$_{3-y}$Cl$_{y}$ (where y is between 0 and 3), and use first principles based density functional theory computations to study their relative formation energies and charge transition levels. We identify vacancy defects and Pb on MA anti-site defect as the lowest energy native defects in each perovskite. We observe that while the low energy defects in all MAPbI$_{3-y}$Br$_{y}$ systems only create shallow transition levels, the Br or Cl vacancy defects in the Cl-containing pervoskites have low energy and form deep levels which become deeper for higher Cl content. Further, we study extrinsic substitution by different elements at the Pb site in MAPbBr$_{3}$, MAPbCl$_{3}$ and the 50-50 mixed halide perovskite, MAPbBr$_{1.5}$Cl$_{1.5}$, and identify some transition metals that create lower energy defects than the dominant intrinsic defects and also create mid-gap charge transition levels.
\end{abstract}

\maketitle{\textbf{ }} \\


The likelihood of defect formation and the positioning of defect levels with respect to band edges are both critically important for applications that are concerned with a semiconductor's optoelectronic characteristics \cite{Def1,Def2,Def3,Def4,Def5,Def6,Def7}. Native point defects could arise as a means of compensation for the presence of impurities, and lead to unintentional conductivity or counteract the prevailing conductivity. Diffusion of impurity atoms in a semiconductor---something that typically happens interstitially---could be assisted by the presence of vacancy defects. Further, ``deep" electronic levels created by low energy defects in the semiconductor band gap adversely affect photovoltaic (PV) performance by potentially causing nonradiative recombination of charge carriers and bringing down efficiencies \cite{Def4,Def5}. It is known from Shockley-Read-Hall theory that defect trap states placed in the middle of the band gap have a very high trapping rate of charge carriers as opposed to shallow defect states \cite{Def_perovs_rev}. On the other hand, such deep levels could also be entangled for quantum sensing or lead to increased absorption of sub-gap energy photons by creating intermediate band PVs \cite{IB1,IB2,IB3,IB4,IB5,IB6,IB7,IB8,IB9}. The experimental determination of the presence, the type and the origin of defects in semiconductors, e.g. via cathodoluminescence (CL) or deep level transient spectroscopy (DLTS), is non-trivial \cite{DefectExpt1,DefectExpt2}. First principles-based density functional theory (DFT) computations provide a useful methodology to study defects, and have been widely applied to accurately predict the defects formation energy and charge transition levels for a large number of crystalline materials \cite{Def2,Corr1,Corr2}. \\

Methylammonium lead halide perovskites with the general formula MAPbX$_3$ (X = I/Br/Cl) have been extensively studied in the last decade or so for optoelectronic applications \cite{HP1,HP2,HP3,HP4,HP5,HP6,HP7,HP8,HP9,HP10,HP11,HP12}. The possibility of doping at MA or Pb sites as well as halogen site mixing provides a large playground for the tuning of electronic structure, absorption coefficients, and defect properties in the MAPbX$_3$ family of perovskites \cite{Comp1,Comp2,Comp3,Comp4,Comp5,Comp6,Comp7,Comp8,Comp9,Comp10}. Recently, we performed a comprehensive study of partial Pb substitution in MAPbBr$_3$ and discovered that a number of transition metals lead to low energy Pb-site defects, are capable of shifting the equilibrium Fermi level and thus the nature of conductivity, and create energy states in the band gap \cite{IB8}. Further, it is seen that mixed halide perovskites have band gaps intermediate to the parent perovskites' band gaps. Mixing of halogen atoms in perovskites has frequently been used to tailor electronic structure, charge transfer and carrier lifetimes \cite{MHP1,MHP2,MHP3,MHP4,MHP5,MHP6}. Partial substitution of I by Br has been shown to enhance the photoinduced halide segregation and charge carrier recombination in MAPbI$_3$, as well as shift certain defect energy levels \cite{MHP7,MHP8,MHP9}.  \\


\begin{figure*}
\includegraphics[width=6.5in]{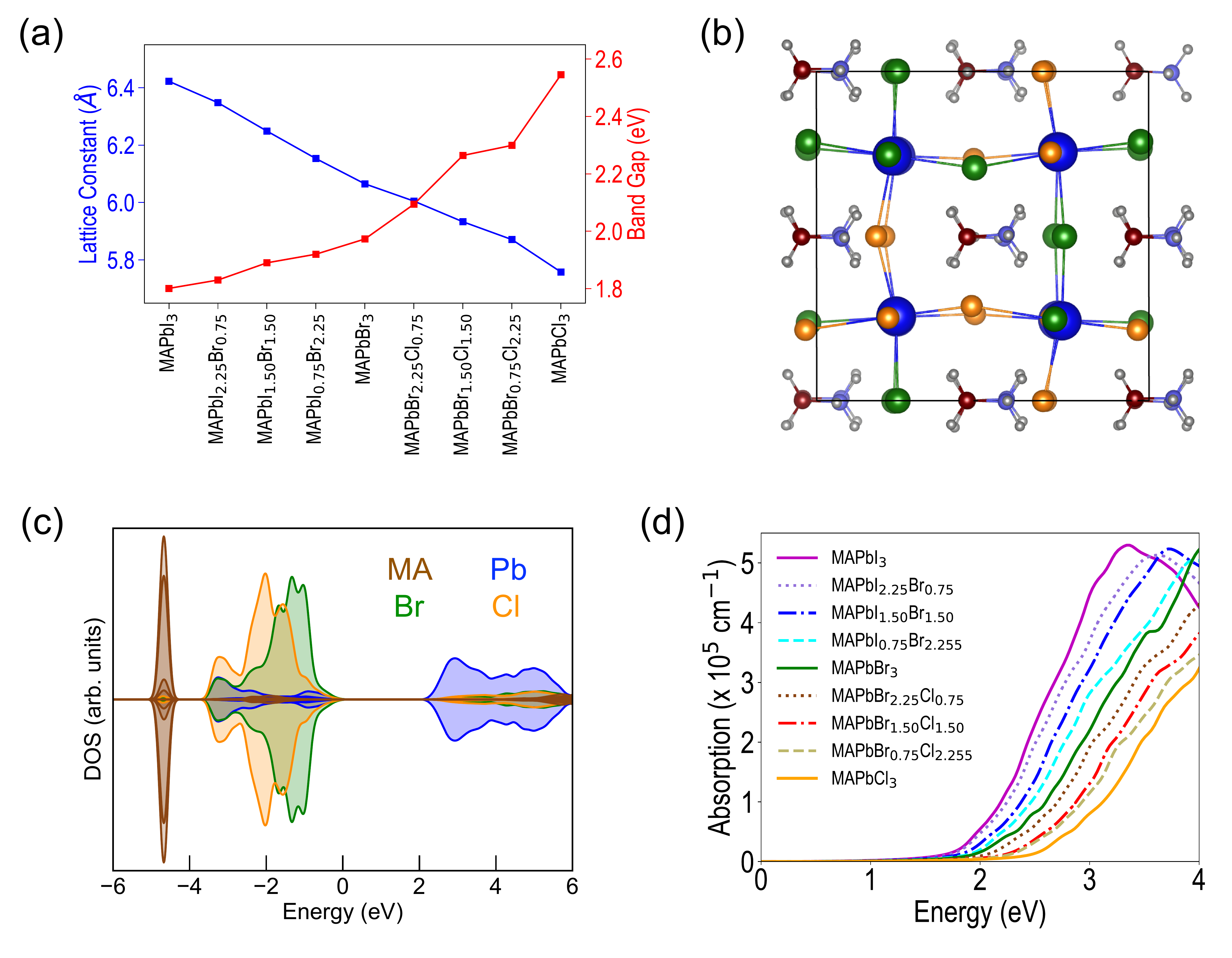}
\caption{\label{Fig:Fig1} (a) DFT computed lattice constant and band gap variation for different perovskite compositions from MAPbI$_3$ to mixed iodide-bromide to mixed bromide-chloride to MAPbCl$_3$. Lattice constant uniformly decreases from iodide to chloride whereas band gap increases. (b) Optimized pseudo-cubic structure of perovskite MAPbBr$_{1.5}$Cl$_{1.5}$ shown as an example. Pb is shown in blue, Br in green, Cl in orange, C in dark red, N in purple and H in silver. (c) Computed density of states for MAPbI$_{1.5}$Br$_{1.5}$. There is uniform mixing between Br and Cl energy states near the valence band maximum. (d) Computed absorption spectra for the 9 perovskites, from MAPbI$_3$ to MAPbCl$_3$. Absorption onset gets delayed moving from iodide to chloride owing to the increase in band gap, but absorption coefficient magnitude follows the trend MAPbI$_3$ $>$ MAPbBr$_3$ $>$ MAPbCl$_3$. (c) and (d) computed using Sumo \cite{Sumo}.}
\end{figure*}


It has been reported in the past that both MAPbI$_3$ and MAPbBr$_3$, the two most commonly used halide perovskites, have impressively high defect tolerance \cite{Defect_tolerance_1,Defect_tolerance_2,HP10}, and owe their utility as PV semiconductors to their optoelectronically benign defects. However, composition engineering at the halogen site can lead to new perovskites where the same defects are no longer benign. Not only can certain defects and impurities become more energetically favorable in mixed halide perovskites (as opposed to pure halides), they could create deeper energy levels and thus have a notable influence on the optoelectronic behavior. Mixed iodide-bromide or bromide-chloride perovskites could be preferred for certain applications due to thermodynamic reasons, availability of precursors, band gap engineering, or other experimental concerns. It is vital to understand the behavior of prominent point defects in such compounds, as well as to screen for possible impurities that could be present in the material or indeed, be incorporated (e.g., at the Pb-site) to affect a certain change in the equilibrium conductivity. \\

In this work, we simulate the structures of several mixed iodide-bromide and bromide-chloride hybrid perovskites and apply density functional theory (DFT) computations to obtain a clear picture of the defect physics in each. We compare the electronic structure and defect properties, specifically the energetics and charge transition levels of dominant point defects, in 9 perovskite systems---MAPbI$_3$, MAPbBr$_3$, MAPbCl$_3$, MAPbI$_{3-y}$Br$_{y}$ and MAPbBr$_{3-y}$Cl$_{y}$ (y = 2.25, 1.50, 0.75). Under ambient conditions, MAPbBr$_3$ and MAPbCl$_3$ are expected to crystallize in the cubic phase whereas MAPbI$_3$ adopts the tetragonal phase \cite{Expt1,Expt2,Expt3,Expt4,Expt5,Expt6}. However, for simplicity of comparison and for ease of simulation of mixed halide systems, we use the pseudo-cubic crystal structures for each perovskite, with computed lattice constants of a = 5.76 {\AA} (MAPbCl$_3$), a = 6.07 {\AA} (MAPbBr$_3$) and a = 6.42 {\AA} (MAPbI$_3$). The mixed halide perovskites were simulated by generating special quasi-random structures (SQS) \cite{SQS1,SQS2} starting from the 2x2x2 supercells of MAPbI$_3$ and MAPbBr$_3$. The pseudo-cubic lattice constants and band gaps computed at the PBE level of theory for all 9 perovskites are shown in Fig. \ref{Fig:Fig1} (a) and listed in Table SI1 along with measured values from the literature \cite{IB7,Expt1,Expt2,Expt3,Expt4,Expt5,Expt6}. The optimized structure of MAPbBr$_{1.5}$Cl$_{1.5}$ (50-50 solid solution of Br-Cl) is shown in Fig. \ref{Fig:Fig1} (b) as an example. \\

It can be seen that going from MAPbI$_3$ to MAPbCl$_3$, the lattice constant steadily decreases whereas the band gap steadily rises from $\sim$ 1.80 eV to $\sim$ 2.55 eV, values that are higher (lower) than reported experimental results for MAPbI$_3$ (MAPbCl$_3$). The structures of all 9 simulated perovskites are shown in Fig. SI1. The energy of formation of any mixed halide perovskite against decomposition to pure perovskites is computed to be less than 15 meV per formula unit, indicating their robust stability; these computed energies are listed in Table SI1. The mixed I-Cl perovskites have energy of formation > 150 meV per formula unit \cite{IB9}, and were thus not considered for this study. In Fig. \ref{Fig:Fig1} (c), we have plotted the computed density of states (DOS) of MAPbBr$_{1.5}$Cl$_{1.5}$, showing that while the conduction band minimum (CBM) is dominated by Pb energy states, the (mixed) halogen energy states determine the valence band maximum (VBM). The computed DOS of all 9 perovskites are plotted in Fig. SI2. Lastly, Fig. \ref{Fig:Fig1} (d) shows the computed absorption spectra of all 9 perovskites, showing clear increasing and decreasing trends, respectively, in the onset of absorption and magnitude of absorption coefficient (y-axis) on going from MAPbI$_3$ to MAPbCl$_3$. The continuous tunability of lattice, optical and electronic structure properties as a function of composition, combined with the reasonable energetic stability, confirms the mixed halide perovskites as a fertile space for designing materials for tailored electronic properties. \\

All DFT computations were performed using the Vienna ab-initio software package (VASP), applying the generalized gradient approximation (GGA) parametrized by Perdew, Burke and Ernzerhof (PBE) and using the projector-augmented wave (PAW) pseudopotentials. The plane wave energy cut-off was set at 500 eV and all atomic structures were fully relaxed until forces on all atoms were less than 0.05 eV/{\AA}. Every calculation for structure optimization, density of states, and simulation of point defects was performed on a 96 atom 2x2x2 perovskite supercell. While trends in lattice constants, band gaps and defect levels are correctly captured when compared to available experiments, deviations arise because of limitations of using a semilocal functional, as well as the neglect of spin-orbit coupling (SOC). Band gaps, band edges and defect energies can be further corrected by the inclusion of SOC \cite{HP12} and hybrid functionals (e.g. HSE06 \cite{HSE_gap}) or GW corrections \cite{GW}. However, we stick to the PBE functional for ease of computation and because properties and trends computed in the past using the same approximations provided a very good qualitative picture of the electronic and defect properties of hybrid perovskites \cite{IB7,IB8}. Our recent work \cite{IB8} shows that corrections from SOC and HSE06 cancel each other out, resulting in a good comparison of PBE computed (without SOC) band gaps and defect levels with measured quantities. The supercell size and the orientation of the MA molecules are other potential sources of errors that are ignored here. \\

For defect calculations, we simulated vacancy, interstitial and substitutional (self and extrinsic) defects in every 96 atom supercell, and optimized the defect structures in various charge states to calculate defect formation energies as a function of the total charge and the chemical potentials of different species, including the electrons. Equation \ref{eqn-1} is used to calculate the formation energy of any point defect in a perovskite MAPbX$_3$--

\begin{equation}\label{eqn-1}
\begin{multlined}
{E^{f}}(D^{q},E_{F}) = E(D^{q}) - 8{E(MAPbX_{3})} - {\mu_{D}} + qE_{F} + E_{corr}
\end{multlined}
\end{equation}

Here, \textit{E(D$^{q}$)} is the total DFT energy of the defect containing supercell in a charge state q, \textit{E(MAPbX$_{3}$)} is the total DFT energy of one formula unit of the bulk perovskite, $\mu_{D}$ is the chemical potential of the relevant species involved in creating the defect, the Fermi level \textit{E$_F$} is the electron chemical potential referenced to the VBM of bulk MAPbX$_3$, and \textit{E$_{corr}$} is the correction energy term calculated using Freysoldt's correction scheme \cite{Corr1,Corr2} to account for periodic interaction between the charge and its image. Defect charge transition levels are defined as the \textit{E$_F$} values where the defect transitions from one stable charged state to another; such levels would correspond to defect states relative to the semiconductor band edges. Chemical potential values are selected from the calculated ranges of stability for every perovskite based on the formation energies of the perovskite and the halide compounds of Pb and MA; this has been shown for MAPbCl$_3$ and MAPbBr$_3$ in Fig. SI3. Before moving to extrinsic substitutional defects, we simulated all possible intrinsic point defects in each of the 9 perovskite systems, namely vacancy (V$_{Pb}$, V$_{MA}$, V$_{I}$, V$_{Br}$ and V$_{Cl}$), self-interstitial (Pb$_i$, MA$_i$, I$_i$, Br$_i$ and Cl$_i$) and anti-site (Pb$_{MA}$, MA$_{Pb}$, Pb$_{I}$, Pb$_{Br}$, Pb$_{Cl}$, MA$_{I}$, MA$_{Br}$, MA$_{Cl}$, I$_{Pb}$, I$_{MA}$, Br$_{Pb}$, Br$_{MA}$, Cl$_{Pb}$ and Cl$_{MA}$) defects. We calculate the charge and Fermi level dependent formation energy of each defect at Pb-rich, halogen-rich and moderate chemical potential conditions. \\


\begin{figure*}
\includegraphics[width=5.5in]{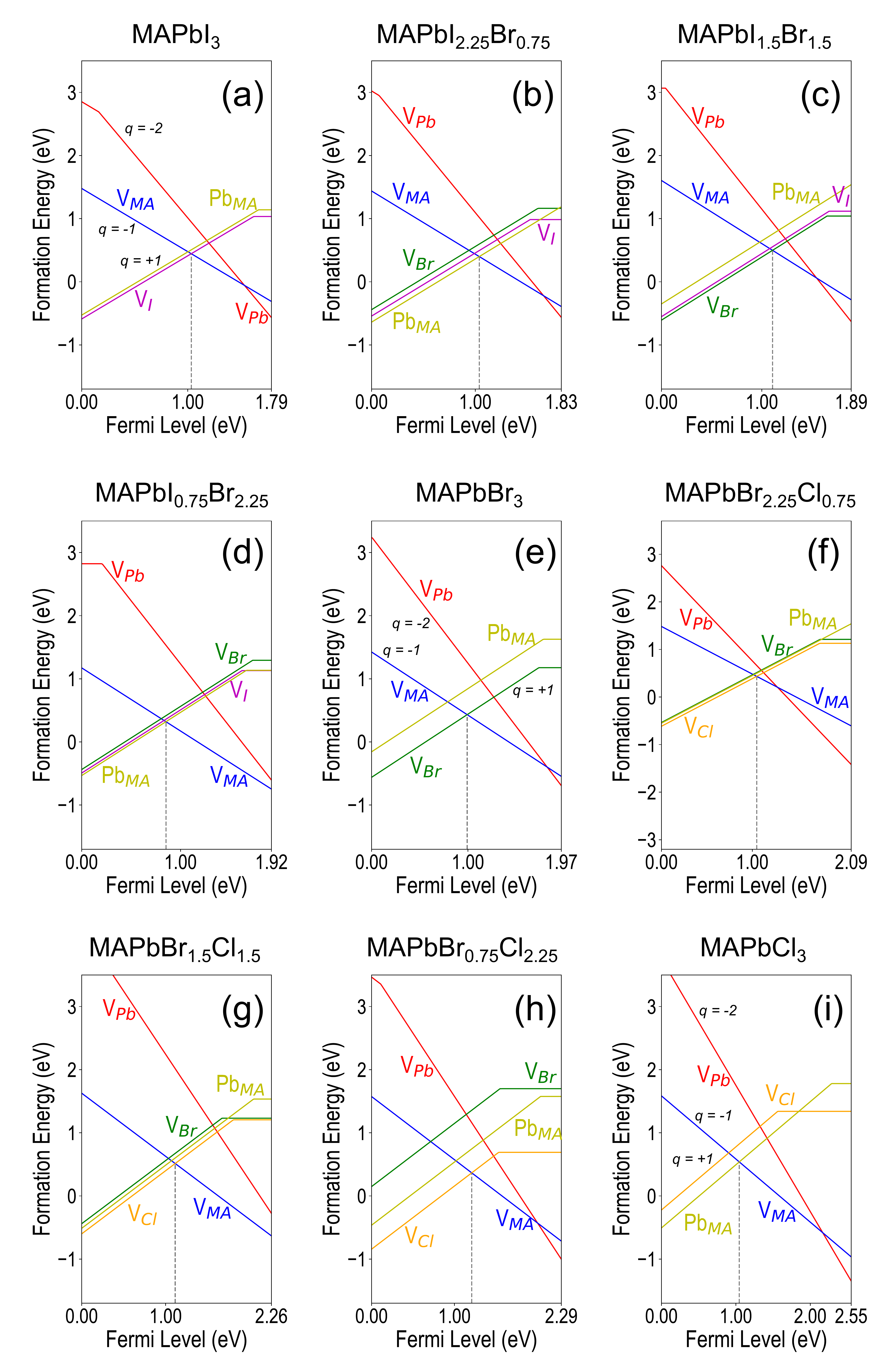}
\caption{\label{Fig:Fig2} Calculated formation energies of intrinsic point defects under Pb-rich chemical potential conditions in (a) MAPbI$_{3}$, (b) MAPbI$_{2.25}$Br$_{0.75}$, (c) MAPbI$_{1.5}$Br$_{1.5}$, (d) MAPbI$_{0.75}$Br$_{2.25}$, (e) MAPbBr$_{3}$, (f) MAPbBr$_{2.25}$Cl$_{0.75}$, (g) MAPbBr$_{1.5}$Cl$_{1.5}$, (h) MAPbBr$_{0.75}$Cl$_{2.25}$ and (i) MAPbCl$_{3}$.}
\end{figure*}


\begin{figure*}[ht]
 \includegraphics[width=6.5in]{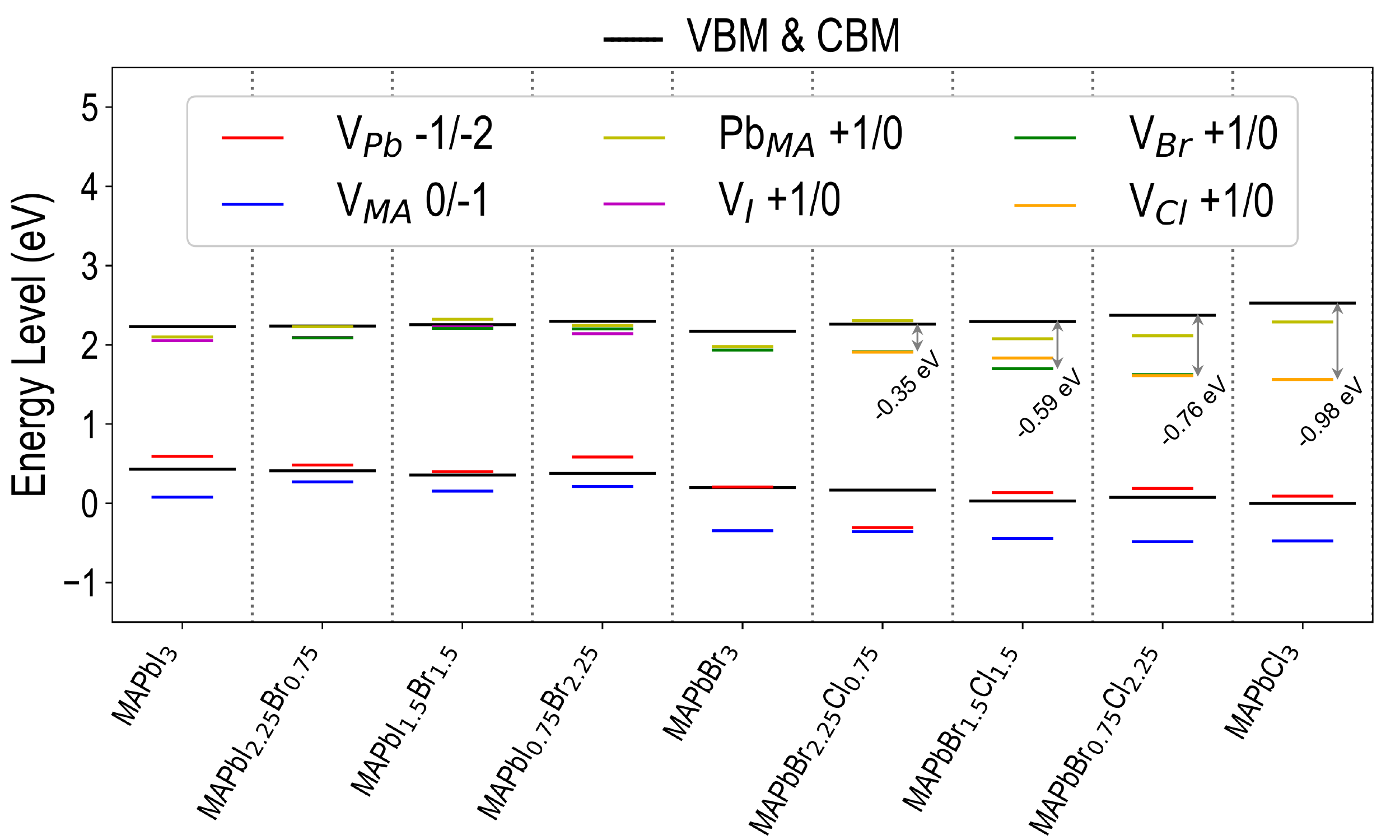}
  \caption{Relevant charge transition levels of dominant intrinsic defects in the 9 perovskites. MAPbI$_{3-y}$Br$_{y}$ compounds show only shallow defect transition levels, whereas the halogen vacancy (V$_{Br}$ and V$_{Cl}$) +1/0 transition level becomes deeper in the band gap for Cl-containing perovskites. Energy levels are aligned using the Nitrogen core 1s state energy from each calculation.}\label{Fig:Fig3}
\end{figure*}

The Fermi level (\textit{E$_F$}) dependent formation energies for the low energy intrinsic defects in the 9 perovskite compounds are plotted for Pb-rich chemical potential conditions in Fig. \ref{Fig:Fig2}, as \textit{E$_F$} moves from VBM to CBM. The complete formation energies for Pb-rich, moderate and halogen-rich chemical potential conditions are shown in Fig. SI4--SI12. The first observation is that each of the vacancy defects, namely V$_{Pb}$, V$_{MA}$ and V$_{X}$ (X=I/Br/Cl), along with the Pb on MA anti-site defect, Pb$_{MA}$, form the set of intrinsic point defects with lowest formation energies in every system. All the remaining anti-site and interstitial defects have higher formation energies, as seen from Fig. SI4--SI12, and are less likely to occur or affect the equilibrium opto-electronic properties of the semiconductor than the defects pictured in Fig. \ref{Fig:Fig2}. 
It is seen that although the acceptor type defects V$_{Pb}$ (q = -2) and V$_{MA}$ (q = -1) and the donor type defects V$_{X}$ (q = +1) and Pb$_{MA}$ (q = +1) all occur in their expected charge states for the majority of the band gap region, there are many cases where these defects show charge transition levels within the band gap. 
We define any transition level that occurs at a Fermi level $>$ 0.2 eV away from the VBM or the CBM as a ``deep" defect level, while the transition levels close to the band edges are termed ``shallow". \\

We observe from Fig. \ref{Fig:Fig2} (a), (b), (c), (d) and (e) that in MAPbI$_3$, MAPbBr$_3$ and each of the iodide-bromide perovskites, the low energy defects create only shallow transition levels, such as the V$_{I}$ or V$_{Br}$ +1/0 transition, or the V$_{Pb}$ -1/-2 transition. While V$_{MA}$ is the lowest energy acceptor type defect, V$_{I}$, V$_{Br}$ and Pb$_{MA}$ all have similar low energies and alternate as the dominant donor type defect. This is consistent with reports that suggest perovskites like MAPbI$_{3}$ have a tendency to release MAI through the formation of vacancy couples, V$_{MA}$ and V$_{I}$, which has negligible impact on the photoluminescence properties \cite{Defect_tolerance_1}.
For every MAPbX$_{3}$ compound, the equilibrium Fermi level \textit{E$_{F}^{eqm}$}, which is determined by charge neutrality conditions \cite{fermi_eqm} and roughly indicated in Fig. \ref{Fig:Fig2} by dashed vertical lines,  
goes from strongly p-type (inside the VB) for halogen-rich conditions to moderately p-type for moderate chemical potential to intrinsic (middle of the band gap) for Pb-rich conditions, as shown in Fig. SI4--SI12 and in Fig. \ref{Fig:Fig2}. The equilibrium conductivity of any perovskite can be tailored by the growth conditions, and different impurities can induce a p-type or n-type shift based on the chemical potential, as will be explained later. \\

In contrast to the MAPbI$_{3-y}$Br$_{y}$ (y = 0, 0.75, 1.5, 2.25, 3) perovskites which don't show deep defect levels, the Cl-containing MAPbBr$_{3-y}$Cl$_{y}$ (y = 0.75, 1.5, 2.25, 3) perovskites show halogen vacancy defect states that go deeper in the band gap with increasing concentration of Cl in the perovskite. As shown in Fig. \ref{Fig:Fig2} (f), (g), (h) and (i), the V$_{Br}$ and V$_{Cl}$ +1/0 transition levels occur at $>$ 0.2 eV from the CBM for each of the Cl-containing perovskites, becoming deeper from (f) to (i). 
This shows that while the halogen vacancies are among the lowest energy donor defects in all perovskite compositions studied here, and \textit{E$_{F}^{eqm}$} follows the same trend from halogen-rich to Pb-rich conditions, deeper intrinsic defect levels are likely to be present in Cl-containing perovskites. This raises the possibility of non-radiative recombination of charge carriers in pure chloride and mixed bromide-chloride perovskites which could lead to lower carrier lifetimes, diffusion lengths and power conversion efficiencies (PCE). While there have been reports of improvement in carrier lifetimes and diffusion lengths upon addition of small amounts of Br or Cl to pure iodide-based solar cells \cite{Lifetimes1,Expt5,Lifetimes2,Lifetimes3}, the highest PCEs are achieved for I-rich perovskites\cite{PCE1,PCE2}. The general lack of success of pure chloride and mixed bromide-chloride perovskite solar cells can be attributed to the low energy deep defects in Cl-rich compositions, and makes these wider band gap semiconductors more suitable for tandem solar cells \cite{tandem1,tandem2} and intermediate band PVs \cite{IB7,IB9}. \\


\begin{table*}[htp]
\caption{\label{table:radii}The elements studied here as extrinsic substituents at the Pb site, the ionic radii for their relevant oxidation states, and their tolerance and octahedral factors in a Bromide and Chloride perovskite lattice.}
\begin{ruledtabular}
\begin{tabular}{ccccccccc}

\textbf{Element}   &    \textbf{Ox. State}    &      \textbf{Ionic Radius (pm)}  &     &  \textbf{t(MAMBr$_3$)}  &  \textbf{$\mu$(MAMBr$_3$)}  &   &   \textbf{t(MAMCl$_3$)} &  \textbf{$\mu$(MAMCl$_3$)}     \\
\hline
\hline
Pb & +2  & 133 & & 0.808 & 0.679 & & 0.813 & 0.735  \\
\hline
Sc & +3 & 88.5 & & 0.935 & 0.452 & & 0.947 & 0.489  \\
\hline
Ti & +3 & 81 & & 0.960 & 0.413 & & 0.974 & 0.448  \\
\hline 
Co & +2 & 88.5 & & 0.935 & 0.452 & & 0.947 & 0.489  \\
\hline
Cu & +1 & 91 & & 0.926 & 0.464 & & 0.938 & 0.503  \\
\hline 
Y & +3 & 104 & & 0.886 & 0.531 & & 0.896 & 0.575  \\
\hline 
Zr & +4 & 86 & & 0.943 & 0.439 & & 0.956 & 0.475  \\
\hline 
Nb & +3 & 86 & & 0.943 & 0.439 & & 0.956 & 0.475  \\
\hline 
Mo & +3 & 83 & & 0.953 & 0.423 & & 0.967 & 0.459  \\
\hline 
Hf & +4 & 85 & & 0.946 & 0.434 & & 0.960 & 0.470  \\
\hline 
In & +3 & 94 & & 0.917 & 0.480 & & 0.928 & 0.519  \\
\hline 
Tl & +3 & 102.5 & & 0.891 & 0.523 & & 0.900 & 0.566  \\
\hline 
Sn & +4 & 83 & & 0.953 & 0.423 & & 0.967 & 0.459  \\
\hline 
Sb & +3 & 90 & & 0.930 & 0.459 & & 0.942 & 0.497  \\
\hline 
Bi & +3 & 117 & & 0.849 & 0.597 & & 0.857 & 0.646  \\

\end{tabular}
\end{ruledtabular}
\end{table*}

In Fig. \ref{Fig:Fig3}, we plotted some of the relevant charge transition levels of low energy defects, namely V$_{Pb}$ (-1/-2), V$_{MA}$ (0/-1), Pb$_{MA}$ (+1/0), V$_{I}$ (+1/0), V$_{Br}$ (+1/0) and V$_{Cl}$ (+1/0), computed across the 9 perovskite compounds. Each transition level as well as the respective perovskite VBM and CBM are plotted alongside each other by referencing all energy levels to the deep nitrogen 1s core state energy from each calculation, with the MAPbCl$_3$ VBM set to energy = 0 eV. 
It can be seen that the V$_{Pb}$ (-1/-2), V$_{MA}$ (0/-1) and Pb$_{MA}$ (+1/0) levels occur close enough to the VBM or CBM to be regarded as shallow defect levels across the 9 perovskites. V$_{I}$ (+1/0) and V$_{Br}$ (+1/0) are also shallow levels in the MAPbI$_{3-y}$Br$_{y}$ (y = 0, 0.75, 1.5, 2.25, 3) perovskites. The V$_{Br}$ (+1/0) and V$_{Cl}$ (+1/0) both become visibly deeper with increasing y in MAPbBr$_{3-y}$Cl$_{y}$ perovskites, going from 0.35 eV below CBM to 0.59 eV to 0.76 eV and to nearly 1 eV in MAPbCl$_3$. To summarize, DFT computations reveal that halogen vacancy defect levels are shallow in iodine containing perovskites but deep in chlorine containing perovskites. \\

\begin{figure*}[htp]
 \includegraphics[width=6in]{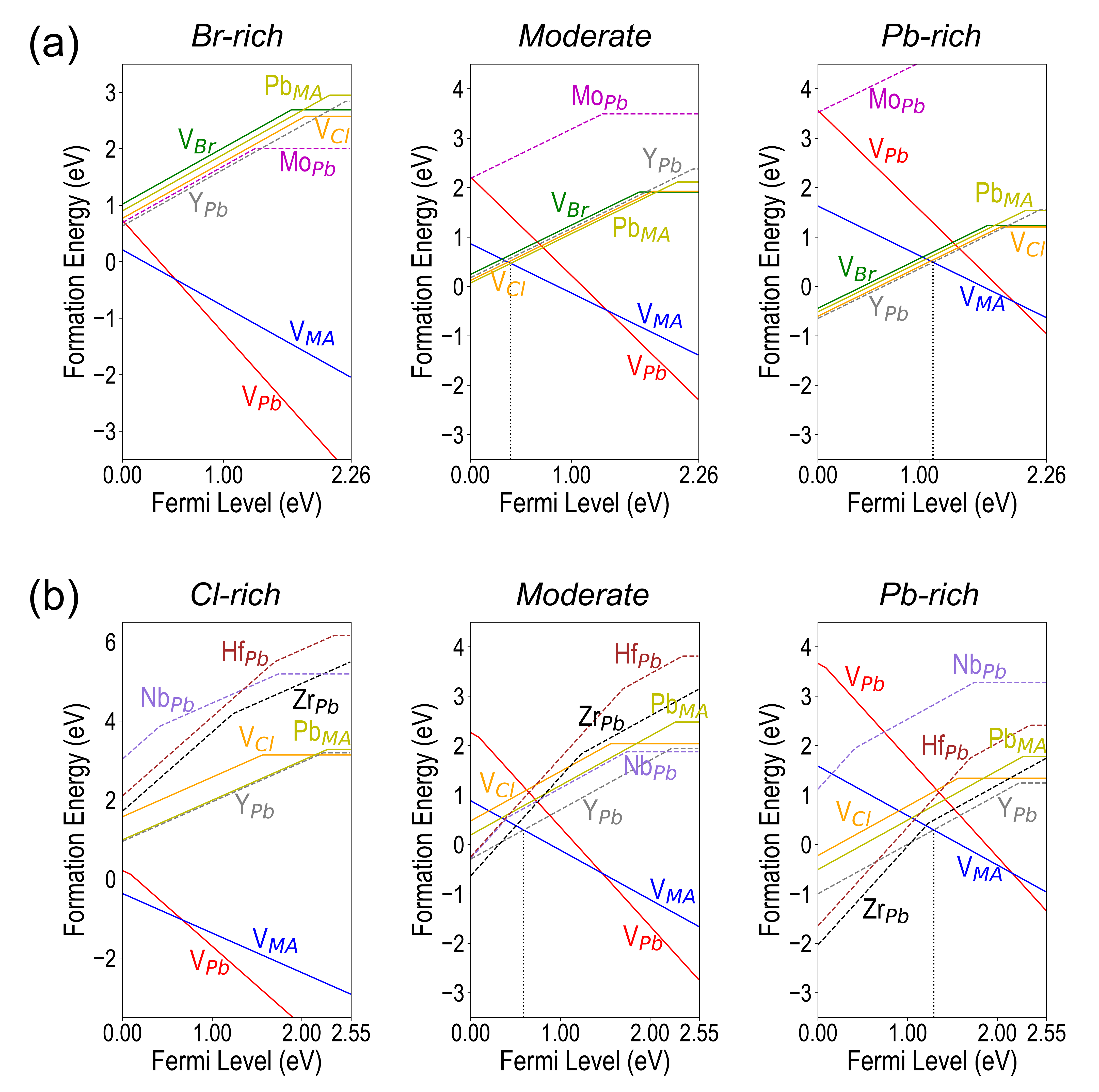}
  \caption{Computed formation energies of extrinsic substitutional defects in (a) MAPbBr$_{1.5}$Cl$_{1.5}$ and (b) MAPbCl$_{3}$.} \label{Fig:Fig4}
\end{figure*}

Impurity atoms or extrinsic point defects can be introduced in a semiconductor to potentially modify the defect or electronic properties as determined by the dominant intrinsic defects. If an extrinsic impurity creates a lower energy acceptor or donor type defect than the lowest energy intrinsic acceptor or donor type defects, respectively, the dominant intrinsic defect(s) can be compensated for, as the equilibrium Fermi level will be changed. We studied this effect by considering several extrinsic substitutional impurities at the Pb site in three perovskites: MAPbBr$_{3}$, MAPbBr$_{1.5}$Cl$_{1.5}$ and MAPbCl$_{3}$, and comparing their computed defect formation energies to the intrinsic defect energetics. In a recently published study \cite{IB8}, we performed an extensive computational study of nearly all period II, III, IV, V and VI elements as substitutional impurities at the Pb site in MAPbBr$_{3}$, and obtained a list of substituents (mainly some transition metals) that create low energy defects and change the equilibrium Fermi level in the perovskite. 
We present the same results here for MAPbBr$_{3}$, 
and extend the study to a set of selected substituents in MAPbBr$_{1.5}$Cl$_{1.5}$ and MAPbCl$_{3}$. \\

In order to determine suitable substitutional impurity atoms for a given perovskite, the Goldschmidt tolerance and octahedral factors \cite{Tol} provide a reasonable estimate of the structural stability of the atom in the perovskite environment. Stable halide perovskites are known to possess tolerance and octahedral factors in the ranges $\sim$ 0.8--1.1 and $\sim$ 0.44--0.9, respectively \cite{IB7}. Using these criteria, we determine a set of suitable substituents that are listed in Table \ref{table:radii}, 
along with their stable oxidation states, ionic radii, and tolerance and octahedral factors in the bromide and chloride perovskite lattices. We performed computations on M$_{Pb}$ substitutional defects in different charged states in MAPbBr$_{3}$, MAPbBr$_{1.5}$Cl$_{1.5}$ and MAPbCl$_{3}$, where M is one of the 15 elements shown in Table \ref{table:radii}. The Fermi level dependent formation energies were computed as before, and we plotted the formation energies for all the low energy extrinsic defects along with the low energy intrinsic defects for halogen-rich, moderate and Pb-rich chemical potential conditions for MAPbBr$_{3}$ in Figure SI13 (reproduced from ref. \cite{IB8}), and  MAPbBr$_{1.5}$Cl$_{1.5}$ and MAPbCl$_{3}$ in Fig. \ref{Fig:Fig4}. \\

It is seen that transition metals have the most interesting behavior as extrinsic impurities in the perovskites, with several of them creating low energy donor type defects that dominate over V$_{Br}$ and V$_{Cl}$. Most of the transition metals prefer oxidation states of +3 or +4, which leads to a net positive charge in the system when they replace the +2 oxidation state Pb atom, explaining why they create donor type defects. Sc, Zr, Nb, Mo and Hf are the low energy impurity atoms in MAPbBr$_{3}$ under different chemical potential conditions, as plotted in Fig. SI13. Fig. \ref{Fig:Fig4} (a) shows that Y and Mo are the only low energy impurities in MAPbBr$_{1.5}$Cl$_{1.5}$ but have comparable energetics to the dominant donor type intrinsic defect Pb$_{MA}$, while from Fig. \ref{Fig:Fig4} (b), it is seen that Y, Zr, Nb and Hf all create low formation energy extrinsic defects in MAPbCl$_{3}$. Each of these stable impurities in each perovskite would move \textit{E$_{F}^{eqm}$} to the right and make the conductivity slightly more n-type; this has been shown in Fig. \ref{Fig:Fig4} (b) for Y$_{Pb}$ in MAPbCl$_{3}$, with very p-type, moderate p-type and intrinsic conductivity resulting in Cl-rich, moderate and Pb-rich conditions, respectively. \\

\begin{figure}[htp]
 \includegraphics[width=\linewidth]{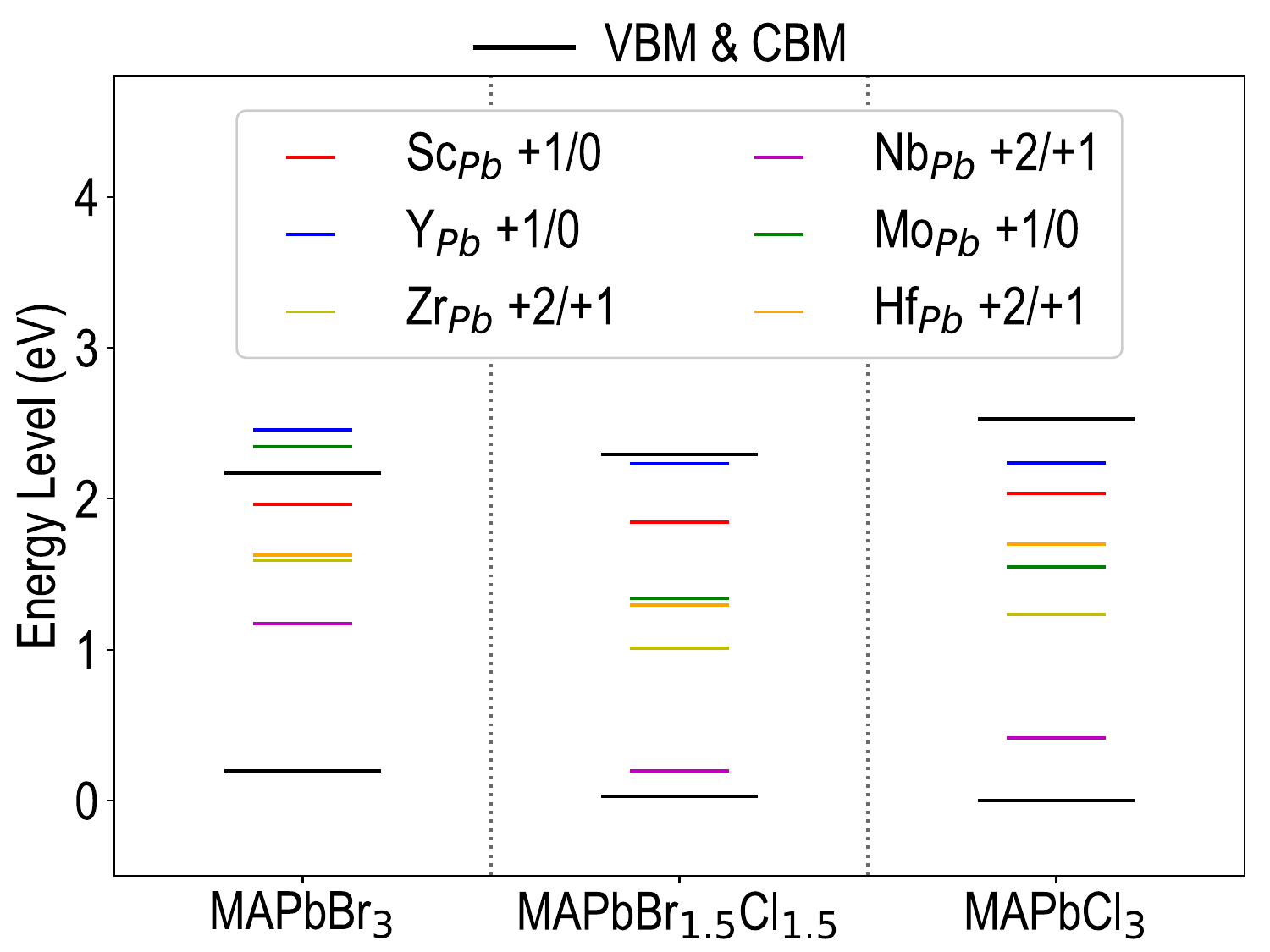}
  \caption{Computed charge transition levels of extrinsic defects in MAPbBr$_{3}$, MAPbBr$_{1.5}$Cl$_{1.5}$ and MAPbCl$_{3}$.} \label{Fig:Fig5}
\end{figure}

It can be concluded that a few transition metals such as Sc, Zr and Hf can potentially create low formation energy donor type defects in bromide, chloride or mixed bromide-chloride perovskites when substituting a Pb atom. However, it should also be noticed that many of these extrinsic defects show deep charge transition levels, owing to the stability of the transition metal in multiple charge states. A comparison between the +2/+1 or +1/0 transition levels of a few selected substitutional impurities are shown in Fig. \ref{Fig:Fig5} for MAPbBr$_{3}$, MAPbBr$_{1.5}$Cl$_{1.5}$ and MAPbCl$_{3}$. Certain levels that are deep in one compound can be pushed closer to the VBM or CBM in others; for example, the Nb$_{Pb}$ +2/+1 level that occurs around the middle of the MAPbBr$_{3}$ band gap is pretty shallow in MAPbBr$_{1.5}$Cl$_{1.5}$ and MAPbCl$_{3}$. The Sc$_{Pb}$ and Mo$_{Pb}$ +1/0 levels, on the other hand, are much deeper in MAPbBr$_{1.5}$Cl$_{1.5}$ and MAPbCl$_{3}$, and will be considered shallow in MAPbBr$_{3}$. \\

Mid-gap impurity transition levels raise the same concern as before of a detrimental effect on PV performance due to non-radiative recombination of carriers. But recent studies by our group and some others have raised the promise of intermediate band photovoltaics functioning via sub-gap photon absorption induced by half-filled impurity energy levels in the band gap \cite{IB7,IB8,IB9}. Such levels can both accept electrons from the valence band and emit them to the conduction band, creating a 2 step photon absorption process which can theoretically increase the solar cell efficiency. The discovery of low energy deep defect creating impurities in halide perovskites thus becomes significant, and could potentially be incorporated and tested in real solar cell materials. We experimentally studied a few such substituents (including Zr) in MAPbBr$_{3}$ as part of our recent work \cite{IB8}, but were unable to conclusively provide evidence of sub-gap absorption. Nevertheless, a complete computational picture of intrinsic and extrinsic defects in halide perovskites as obtained in this work provides us with a list of possible substituents that can be introduced in bromide or mixed bromide-chloride perovskites of various compositions to suitably change the defect and electronic properties for PV applications. \\

In summary, we used first principles computations to study the trends in defect formation energies, defect energy levels and equilibrium Fermi level for various types of point defects in different compositions of pseudo-cubic methylammonium lead halide perovskites. We observed that the lowest energy intrinsic defects, namely the vacancy defects and Pb on MA anti-site defect, only create shallow levels in the iodide, bromide and mixed iodide-bromide perovskites, but create deeper levels in the pure chloride and mixed bromide-chloride perovskites, which could be unfavorable due to the danger of non-radiative recombination of charge carriers. We further studied several extrinsic substituional impurities at the Pb-site in three perovskites, MAPbBr$_{3}$, MAPbBr$_{1.5}$Cl$_{1.5}$ and MAPbCl$_{3}$, and discovered that certain transition metals like Sc, Zr, Hf, Mo and Y can create donor type defects with lower energy than the potentially deep level creating halogen vacancy defects. Such impurities can dominate over the intrinsic defects and change the nature of conductivity in the perovskite, while themselves creating energy levels in the band gap that can potentially be exploited for sub-gap absorption and increased PV efficiencies. \\

This material is based upon work supported by Laboratory Directed Research and Development (LDRD) funding from Argonne National Laboratory, provided by the Director, Office of Science, of the U.S. Department of Energy under Contract No. DE-AC02-06CH11357. Use of the Center for Nanoscale Materials, an Office of Science user facility, was supported by the U.S. Department of Energy, Office of Science, Office of Basic Energy Sciences, under Contract No. DE-AC02-06CH11357. This research used resources of the National Energy Research Scientific Computing Center (NERSC), a DOE Office of Science User Facility supported by the Office of Science of the U.S. Department of Energy under Contract No. DE-AC02-05CH11231. We gratefully acknowledge the computing resources provided on Blues and Bebop, high-performance computing clusters operated by the Laboratory Computing Resource Center (LCRC) at Argonne National Laboratory. \\

\section*{References}
\bibliography{aipsamp}

\end{document}